# Organic Primitives:
# Synthesis and Design of pH-Reactive Materials using Molecular I/O for Sensing, Actuation, and Interaction

**Viirj Kan**[1], **Emma Vargo**[1], **Noa Machover**[1], **Hiroshi Ishii**[1],
**Serena Pan**[1], **Weixuan Chen**[1], **Yasuaki Kakehi**[1,2]
[1]MIT Media Lab, Cambridge, USA  [2]Keio University, Fujisawa, Japan
Viirj@media.mit.edu, {evargo, noamori}@mit.edu, ishii@media.mit.edu, pan27@mit.edu,
cvx@media.mit.edu, ykakehi@sfc.keio.ac.jp

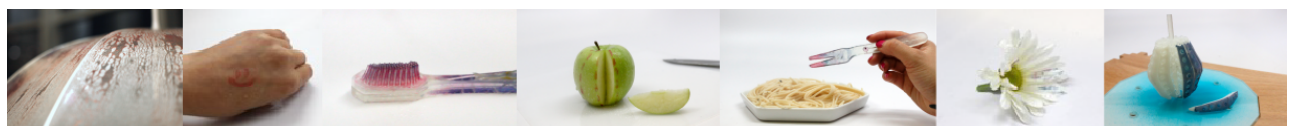

**Figure 1.** Application prototypes and scenarios using *Organic Primitives*: (left to right) acid rain sensing umbrella; color changing cosmetics; saliva sensing toothbrush; apple as sensor and display; flavor augmenting utensil; flower tattoo; bionic fruit.

## ABSTRACT
In this paper we present *Organic Primitives*, an enabling toolbox that expands upon the library of input-output devices in HCI and facilitates the design of interactions with organic, fluid-based systems. We formulated color, odor and shape changing material primitives which act as sensor-actuators that convert pH signals into human-readable outputs. Food-grade organic molecules anthocyanin, vanillin, and chitosan were employed as dopants to synthesize materials which output a spectrum of colors, degrees of shape deformation, and switch between odorous and non-odorous states. We evaluated the individual output properties of our sensor-actuators to assess the rate, range, and reversibility of the changes as a function of pH 2-10. We present a design space with techniques for enhancing the functionality of the material primitives, and offer passive and computational methods for controlling the material interfaces. Finally, we explore applications enabled by *Organic Primitives* under four contexts: environmental, cosmetic, edible, and interspecies.

## ACM Classification Keywords
H.5.m Information interfaces and presentation (e.g., HCI): Miscellaneous

## INTRODUCTION
Whether it be the information embodied within the rain, the ocean, a dinner plate, or in human tears, the flow of information through fluids provides us a glimpse into the biological and chemical states of systems. A large part of our everyday experience with natural systems is inaccessible to HCI designers and engineers for user interaction design. The challenge of programming interactive experience within the realm of food, personal care, and other organic systems lies largely in the limitations of conventional tools in HCI, as they are incompatible with these soft, fluid-based systems. The dialects required to manipulate such systems are mediated by chemical codes, which control the most complex machine we know —life. In order to open up possibilities for interacting with organic systems in HCI, an accessible, biocompatible toolbox and design process is essential. This paper offers a set of combinable material-based sensor-actuators termed *Organic Primitives*, enabling non-Chemists to design for fluid inputs.

The toolbox consists of color, odor and shape changing material primitives that sense contents within fluids and convert them into sensory information. The material primitives can output a spectrum of colors, different degrees of shape deformation, and switch between odorous and non-odorous states based on pH value. The scope of this paper focuses on sensing pH in fluid as a starting point, with the goal of developing a model methodology for future sensor-actuator development. The motivation for this toolbox is to enable designers without extensive chemistry backgrounds to create a variety of multi-modal composites by simply selecting and combining responsivities of interest. The design of our toolbox addresses five key criteria: 1) Solid-State: liquid state pH-reactive phenomena of the molecules must be synthesized into solid state materials; 2) Flexibility: materials should have the ability to be synthesized into a range of form factors, without drastically differing in their fabrication processes; 3) Accessibility: they should be safe for handling, producible in a kitchen environ-ment and utilize reagents that are easy to attain; 4) Human-Readable: output properties should be within receptive fields and thresholds of human perception; 5) Combinability: material primitives should be compatible with one another to yield multi-property inheritances. In this paper, we demonstrate how we address these criterias through the development of our material primitives.

The sensor-actuators were synthesized by doping select biopolymers with organic molecules which offer optimal flexibility for synthesis across a range of forms. Food-grade organic compounds anthocyanin, vanillin, and chitosan were employed as accessible dopants to formulate materials which output a spectrum of colors, degrees of shape deformation, and a switch between odorous and non-odorous states. Their molecular reaction to pH enables pH information to be both sensed and communicated to the user. We present methods for how we bring these liquid-state pH reaction mechanisms into solid-state material primitives in order to be used as sensor-actuators for HCI. We evaluated the individual output properties of our primitives within this toolbox to exhibit the human-readable rate, range, and reversibility of the changes as a function of pH 2 to 10. The activation and control section exhibits how computational systems can be integrated with the material primitives. We demonstrate the primitives are self-compatible and can be combined with one another to form multi-inherited property outputs in the Design Implications section. We showcase techniques for how the primitives can transcend beyond mere input-output devices to achieve higher order complexity in the Design Space. Finally, we present Molecular Design Interactions in four application contexts to demonstrate the possibilities enabled by this toolbox for human interaction with the environment, body, food and interspecies (figure 1).

## RELATED WORK

### Interfacing with Biological and Chemical systems

Biocompatible materials for sensing and actuation are becoming increasingly relevant for the HCI community with the rise of research towards soft, low power devices that operate close to and within the human body[20]. Recent developments like bionic contact lenses leverage information excreted from the body such as the composition of tears, for medical diagnosis [13]. Ingestible devices such as Google X's pill-based authentication use stomach acids to power the biodegradable device as a temporary key for unlocking devices [32]. These research contributions typically require a background in chemistry or biology. There have been efforts to increase accessibility of synthetic biology through standardized DNA building blocks called BioBricks and miniaturized lab equipment such as the Amino[33][24]. However, genetically engineering organisms for sensing and actuation currently remains difficult in practice, as methods are often irreproducible due to variability across organisms[36][4][17]. To advance the interests of HCI and facilitate human interaction with biological and chemical systems, an accessible toolbox of input-output devices can be achieved through molecules synthesized into material primitives.

### Molecules for Sensing and Actuation

The complexity of biomolecules makes detection using modern electronic chemical sensors a challenge. Researchers in chemical and biological engineering fields often rely on chemical mechanisms and reactions for sensing and analysis. Bicinchoninic acid assay is an analytical procedure utilizing molecules copper(II) sulfate pentahydrate and bicinchoninic acid to form reactions to determine the concentration of proteins in solution[37]. The physical nature of a compound's composition can lend itself to complex reactions, offering capabilities to sense and move based on physical property gradients. Water droplets mixed with food dye have exhibited sensing and motility behaviors, termed *artificial chemotaxis*, due to the relative evaporation rates and surface tension of two-component fluids. Researchers have exploited these properties to create autonomous fluidic machines [8]. While there has been little emphasis within these fields to develop input-output devices that cater to the interests of HCI for interface design, these molecular techniques for chemical sensing and analysis can be leveraged.

### Material-Mediated Interface Design

*"The most profound technologies are those that disappear. They weave themselves into the fabric of everyday life until they are indistinguishable from it." - Mark Weiser*[40]

Recent developments in HCI and robotics have sought to achieve seamless technological devices by incorporating sensing, information processing, and actuation within a unitary material[21] [27]. The development of wearable technologies that are compact, mobile, and conform to the body demonstrate a growing value in material and molecular-based interfaces [9] [25] [35] [39]. Researchers have investigated soft, compliant methods such as thin-film electronics in Unimorph, silicone rubber in the Harvard Soft Robotics Toolkit and bacterial spores in bioLogic to create compact interactions [19][1][45]. The notion of interfaces which can deliver color change [22], odor release [23] and shape transformation [30] have been individually explored. However, compact interfaces with sensing and multiple output functionality remains a challenge due to limited tools in HCI. *Organic Primitives* offers an adaptable, molecular scale method to facilitate HCI researchers in exploring such design possibilities.

## DESIGNING PH-REACTIVE MATERIAL PRIMITIVES

To enable user interaction with organic systems, we propose utilizing the sensing-actuation capabilities of food-grade flavor molecules and edible materials, by employing them as molecular I/O mechanisms to build input-output devices for HCI. These molecules are encountered daily in food, but their stimuli-responsive properties are generally overlooked. Such renewable sensor-actuators are inexpensive to produce, safe to ingest, and offer complex multi-modal capabilities. They can be used as independent material sensor-actuators or supplemented by integrating with traditional electronic devices.

pH value is an intrinsic attribute among all fluids and serves as an important indicator in a broad range of systems, from nanoscale microbial systems to planetary-scale marine ecosystems. For this reason, we begin our material sensor-actuator development with pH-reactive primitives. Chemically, pH is the measure of the acidity and alkalinity with the relative amount of free hydrogen and hydroxyl ions in a fluid. It is an important health indicator for many complex chemical and biological processes in, out, and on the human body, from food to personal care. pH also serves as a critical environmental indicator for processes such as ocean acidification and atmospheric contamination. Many organic molecules respond to pH, but only a subset of these are edible and yield human-readable changes like color, odor, shape, and taste. Our material-based sensor-actuators begin with the selection of dopants and biopolymer substrates. Within our approach, we utilize the Brønsted-Lowry acid-base reaction as a core driver of our material transformations due to the rapid, reversible, and bidirectional nature of the reaction. In this section, we introduce pH-reactive organic molecules anthocyanin, vanillin and chitosan as color, shape and odor changing dopants, along with kappa-carageenan and sodium alginate as biopolymer substrates for material synthesis.

**Molecular Dopant Selection**

In this section, we discuss the chemistry of the pH-reactive compounds. These molecules were selected for their safety, accessibility, and biocompatibility, so researchers can develop their own pH-responsive material interfaces without special tools or facilities.

*Anthocyanin: Color-Changing Reconfigurable Molecule*
All tissues of vascular plants contain the flavonoid anthocyanin, a pigment that changes color under varying pH solutions. In the 1664 book *Experiments and Considerations Touching Colours* by chemist Robert Boyle, various edible plants are reported as visual pH indicators due to pH-responsive mechanisms in their tissues [6]. Anthocyanin is commonly used as a food colorant in the food and beverage industry, and has been found to possess anti-inflammatory antioxidant properties[18]. It has also been researched for use as an indicator for packaging applications to detect spoilage in pork and fish products[46].

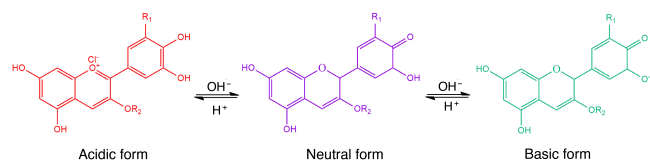

Figure 2. Chemical diagram of color-changing anthocyanin pH reaction

Under different pH conditions, the hydroxyl (OH) and/or methyl ether (O-$CH_3$) groups attached to the carbon rings (figure 2) undergo reversible structural transformations and ionizations. Restructuring a molecule changes the way it absorbs light, giving rise to color changes [3] [31].

*Vanillin: Odor-Switching Flavor Molecule*
Vanillin is a flavor molecule found in vanilla beans, contributing to their characteristic aroma. The smell of vanillin has been found to elicit feelings of relaxation [38], offering potential uses for interface design. The effect of pH on vanillin stability was first characterized by food scientists in the early 1970s [42]. In 1990, Flair and Setzer determined vanillin could be used as an olfactory titration indicator for blind students [15]. Vanillin is prevalent as a flavoring for food products such as chocolate and ice-cream[5].

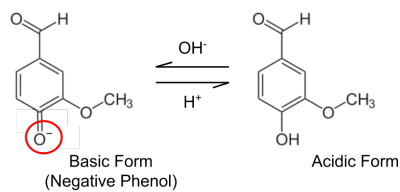

Figure 3. Chemical diagram of vanillin pH reaction

The smell of vanillin is elicited at pH 2-8 and suppressed at pH 9 and 10. We hypothesize that the elimination of odor at high pH is due to vanillin's phenol group pKa value of 7.38, such that it becomes negatively charged, and therefore less volatile, under basic conditions (figure 3) [10].

*Chitosan: Swelling and Shape-Memory Macromolecule*
Chitosan is a polysaccharide derived from the exoskeletons of shrimp and crustaceans, widely studied due to its abundant and biocompatible nature [12] [29] [16]. The pH-responsive swelling of chitosan hydrogels was first characterized in publications from the 1980s [44]. It has been researched for drug delivery applications, such as to target tumor tissues through the release of drugs at specific pH[2]. The tunability and shape memory characteristics of chitosan have been studied in tissue engineering to create scaffolds with impregnated bioactive agents for guiding cellular growth towards generation of new tissue[26].

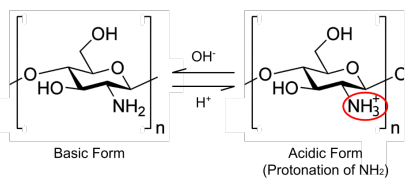

Figure 4. Chemical diagram of chitosan pH reaction

Illustrated in figure 4, its $NH_2$ amino functional group is protonated in low pH conditions, causing the polymer to swell due to charge repulsion; if the pH becomes high, chitosan returns to its collapsed state.

**Material Synthesis**

In order to utilize the pH-reactive properties of the molecules, the chemical phenomenon must be materialized from a liquid into a solid state. We have found sodium alginate and kappa-carrageenan to be good biopolymer substrates for the pH-responsive molecules due to their semi-permeable structures, consistent behavior, rapid diffusion rates, and relatively neutral pH. Primitive subsets with different range, rate and reversibility of change can be generated by tuning synthesis parameters such as cross-linkers, co-polymers and concentrations. Kappa-carrageenan was used as the base material for our anthocyanin color-changing films. Electrostatic interactions between the two molecules allow them to create a stable compound; kappa-carrageenan is an anionic polysaccharide, while anthocyanin is cationic[14]. To synthesize an odor-changing film, vanillin was encapsulated in an alginate gel to form our scent-changing solution. Sodium alginate was cross-linked with calcium sulfate right before mixing, creating a tangled matrix to hold the vanillin molecules. Chitosan is a structural material and does not require a matrix. We synthesized shape-changing films by dissolving dried chitosan powder in 3% acetic acid and drying it in sheets. Higher concentrations of chitosan result in thicker, stiffer sheets. Concentrations of the three materials and dopants should be tuned for the application; an example composition of each material is given in the next section.

**OUTPUT CHARACTERIZATION**

In order to evaluate the human-readability and assess the design parameters of our color, odor and shape changing sensor-actuators, we characterized each primitive's range, rate and reversibility as a function of pH value. pH solutions were made using food-grade materials to ensure safety and accessibility: citric acid ($H_3Cit$), an acid derived from citrus fruits; sodium citrate (NaHCitrate), a common food additive; sodium bicarbonate (NaHCO), also known as baking soda; sodium hydroxide (NaOH) or lye, commonly used for food preparation; and deionized water ($H_2O$). The concentrations we used comply with the Federal Food and Drug Administration's (FDA) Generally Recognized as Safe (GRAS) listing. The solutions were tested and verified using the Oakton pH5+ EW-35613-52 pH meter, as well as pHydrion litmus strips.

**Color as a Function of pH**

Our color-changing samples were prepared using 1.5%w/v kappa-carrageenan in deionized water doped with 0.1% w/v anthocyanin, purchased from Modernist Pantry and Enasco respectively. We tested samples with our pH 2-10 solutions. The samples were sprayed uniformly with pH solution, allowed to dry, and an X-Rite CMUNPH spectrophotometer was used to find a CIE L*a*b* color value for each sample.

*Range and Spectrum of Color Change*

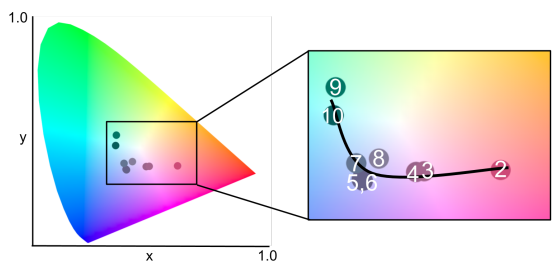

**Figure 5. CIE L*A*B* color space of color change ranges of Organic Primitive sample made with 1.5%w/v kappa-carrageenan doped with 0.1%w/v anthocyanin.**

Depending on pH, the sample color ranges from redder tones to green (figure 5). Materializing anthocyanin into a solid-state film with our selected biopolymer does not negatively affect its responsiveness.

*Rate of Color Change*

The speed of the color change was determined by filming the color primitives and applying pH 2-10. The results were qualitatively assessed as it is difficult to define specific start and stop times. Color noticeably begins to change within 11 milliseconds after solution is applied. They reached their final color between 50 to 90 seconds. Acidic pH's 2-4 are appear to be more rapid than 7-10. pH 5 and 6 were the slowest amongst all the samples. This is consistent with the different ion concentrations in the test solutions.

*Reversibility of Color Change*

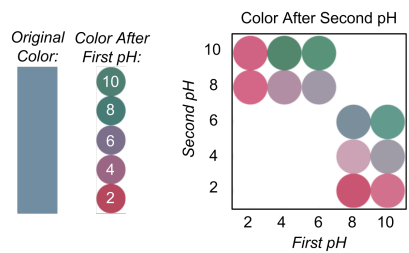

**Figure 6. The reversible effects of layering pH.**

In this study, the samples were sprayed with the first pH solution, given an hour to equilibrate and then documented. The dry sample was then sprayed with the second pH solution. After an hour, RGB values were extracted from digital photographs taken under a lighting hood. Results from figure 6 show that only certain pH responses can be reversed. This is due to the constraints of the substrate biopolymer contained in this primitive, as anthocyanin color changes in fluid are entirely reversible reactions. We hypothesize that this incomplete reversibility is due to slower diffusion rates in solid materials and buffering effects of the polymer.

**Odor as a function of pH**

Our odor-changing samples were prepared by dissolving 1 g vanillin (Sigma Aldrich) in 10 mL 200-proof alcohol to create a vanillin stock solution. We then made a solution of 0.12% v/v vanillin stock and 1.5% w/v sodium alginate in deionized water. Immediately before pouring, the sodium alginate was cross-linked by adding 15 parts calcium sulfate to 100 parts vanillin-sodium alginate solution. Samples were tested with the same pH 2-10 solutions as previous trials. To evaluate the human perception of our odor-changing films, we conducted a user study with 14 subjects, 4 male and 10 female. We had participants rearrange nine 1" x 1" samples of pH-activated odor material from pH 2-10, based on odor intensity.

*Range of Odor Strengths*

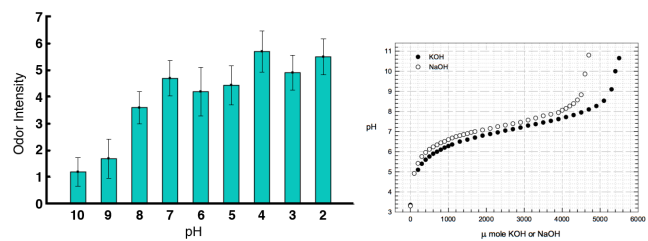

**Figure 7. Odor rearrangement results (left) where error bar represents standard error. Titration curve (right) by Frenkel, showing vanillin molecule response to pH.**

In figure 7, our results show that users can clearly distinguish the on-off switching behavior of the odor material based on pH. pH 9 and 10 were ranked as odors with no smell or low intensity, whereas samples with low pH generated a stronger odor perception. The user's sample arrangements for odor intensity correlated with Frenkel's titration curve and response of vanillin molecule to pH [7].

*Rate of Odor Change*

Two independent researchers qualitatively assessed the rates of odor change using a timer. Vanillin films as prepared above were cut into nine 1" x 1" samples and 640uL of each pH solution 2-10 were applied to a film. pH 2-8 exhibited a sweet odor released within 1-5 seconds of depositing solution. After 30 or more seconds, the odor intensified over the course of hours and maintained its smell after dried. With pH 9-10, approximately 30 seconds after a pH solution is deposited, users were able to establish that the material had no smell. Within the first 30 seconds, the material possessed a faint vanillin smell. Our preliminary study exhibits that each applied pH results in a different time-intensity relationship. This phenomena will be better examined in future studies, using olfactometry test methods as defined by the American Society of Testing and Materials ASTM E679 and E544 [28] and quantitative analysis with headspace gas chromatography.

*Reversibility of Odor Change*

Based on qualitative assessments, we found that our odor materials could be activated from a low pH (smell on) to a high pH (smell off). However, after a high pH change, the material was unable to reactivate in its smell. We hypothesize that due to the low concentration of vanillin in our materials, each time it is activated by pH solution the odor molecules are consumed by evaporation into the air.

## Shape as a Function of pH

Chitosan samples were prepared with 4% w/v chitosan powder (Spectrum) dissolved in 3% v/v acetic acid (Sigma Aldrich). After drying, the film was cut into 20 x 7.5 mm strips, and tested using pH solutions of 2, 4, 6, 8, and 10. For each pH, five 0.75 uL droplets of the pH solution were applied across the sample's 7.5 mm centerline with a micropipette, forming a hinge. The sample bent along this line over a period of approximately 10 minutes. Samples were filmed and a still image for each sample was taken from the footage 4, 7, and 10 minutes after the droplets were applied. The software ImageJ was used to quantify the bend angle of the hinge. Results are the average of two identical trials.

*Range of Angle Change*

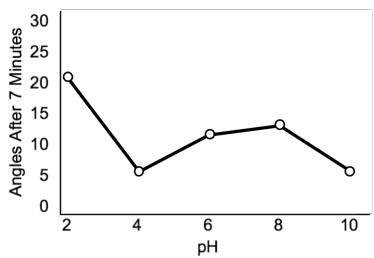

**Figure 8. Angle change 7 minutes after droplets are applied to a 4% w/v chitosan film. The bimodal trend matches literature results [ 6].**

Our results as shown in figure 8 reveal a bimodal relationship between bend angle and pH within the range of pH 2-10. Mahdavinia et al. found a similar relationship in their study, with swelling maxima at pH 3 and 8. Decreased swelling at middle pHs may be attributed to cross-linking [ 6].

*Rate of Angle Change*

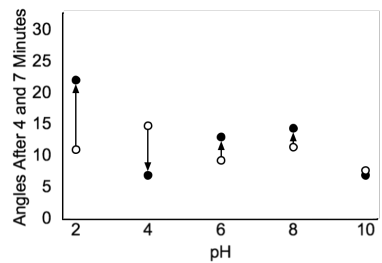

**Figure 9. Changes in bend angle for 4% w/v chitosan films between 4 and 7 minutes after pH solution is added. Upward arrows indicate an increase in bend angle.**

Regarding time evolution, the two pHs corresponding to angle minima, 4 and 10, decreased in angle between 4 and 7 minute measurements (figure 9). The other pH solutions had an increase in angle. It may be that interaction with any solution causes the chitosan to swell approximately the same amount, and then after some time the swelling becomes pH-dependent.

*Reversibility of Angle Change*

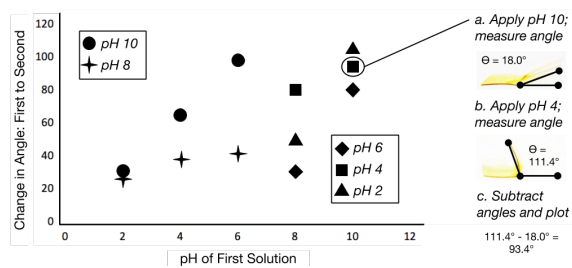

**Figure 10. Reversibility graph of layering different solutions.**

For this study, five 0.75 uL drops of the first pH solution were applied across the center of the sample. The film was given an hour to reach its final angle, measured with ImageJ (a). After the hour, droplets were applied again with a second pH solution on top of the first solution(b). Figure 10 shows that the more the second solution differs in strength from the first, the greater the change in angle is going to be. This preliminary study is promising for shape reversibility, as the bend angle of the film can be predictably manipulated by applying high and low pH solutions on the same material.

## ACTIVATION AND CONTROL

Material activation generally fall into one of three categories of control—global, local or discrete, as shown in image a, b, and c of figure 11. As shown in figure 11, global control utilizing atomizers and nozzles can activate a large area of the material. Local control can be developed by fabricating pipes and channels into the material. Discrete control can be implemented by designing specified outlets in fluidic circuits. Computational control can enable temporal control of a primitive's output by sequentially defining specific pH values deposited.

## Coupling and Decoupling pH Inputs and Outputs

Fluid inputs from natural contexts as well as fabricated solutions can be used to activate material outputs. Natural inputs include rain drops, ocean water, body fluids, food sources, and organismic excretions from plants and microbes. Fabricated inputs using prepared acid-base solutions can offer designers a wider output range while yielding more precise outputs. Fabricated pH solutions can be integrated with mechanical and electronic systems to develop machine-mediated inputs, as shown in image d of figure 11.

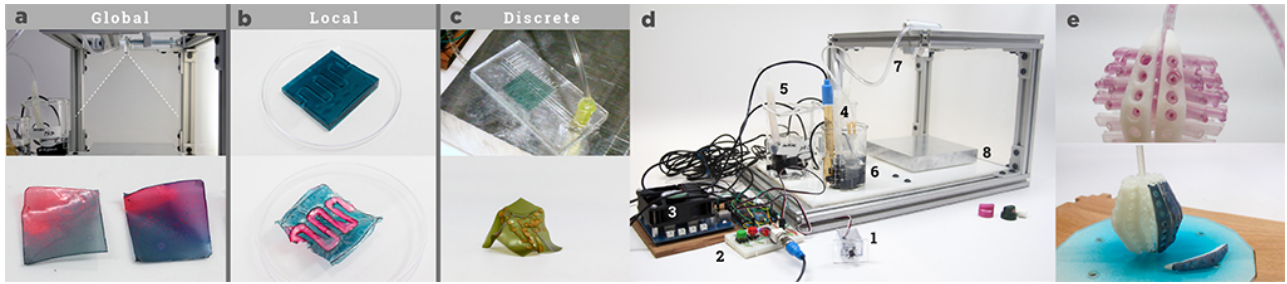

**Figure 11. Techniques for activation and control of the material primitives:** a) Global activation involves pH solution sprayed onto the entirety of a material in an instance b) Local activation actuates a stream of solution in a channel c) Discrete activation triggers a single droplet in a precise position to actuate d) Hydraulic system for global and sequential activation e) Bionic fruit device deposits material primitives for use as edible films.

**Computational Activation**

While it is not required in order to achieve an input-output response, electromechanical systems can be integrated for activation (figure 11-d) or deposition(e). The apparatus in image d is a platform which computationally generates a desired pH solution for global activation of an *Organic Primitive* as shown in image a. It can be used to control the outputs in a particular sequence, to enable the material to animate and perform a series of behaviors, similar to figure 15. It possesses 2 reservoirs of solution - pH 2 and pH 10, and outputs any desired pH these values. Using a series of hydraulic pumps and a pH sensor to complete the closed loop control of the system, we can continually adjust the pH value as we measure with a digital pH meter. A microcontroller with a potentiometer enables a user to select the desired pH by rotating the knob(d-4), initiating the pumps to draw fluids from pH 2 and pH 10 reservoirs(d-5) into a microfluidic mixer(d-4). Once the desired pH value is reached, the user can deposit the pH solution through the nozzle (d-7) to activate a material on the platform (d-8). The nozzles we used were adopted from off-the-shelf spray paint tops, offering a low-cost and versatile device. For deposition, image e shows a bionic fruit which deposits primitives on the surface of a 3D printed fruit for user interaction. Further elaboration is in figure 26.

Local activation (image b of figure 11) is created by casting hollow channels with the material primitives. A number of fluidic logic can be implemented to transcend the capabilities of the primitives from mere sensor-actuators, as illustrated in figure 18 and 19. Discrete activation (image c of figure 11) can be achieved by adapting our apparatus in image d to connect to a microfluidic device instead of the nozzle. The microfluidic device was fabricated from acrylic with a lasercutter to achieve 600 micron channels. It contains outlets which deposit pH solution to the material in small droplets. By depositing discrete droplets of pH solution onto a shape-color changing material, it can output a complex structure, as shown in image C of figure 11.

**DESIGN IMPLICATIONS**

A limitation in these pH-reactive primitives is that they require liquid to be actuated. This presents novel design constraints, such as the ability to utilize natural inputs like rain water or prepared fluid inputs that are computationally activated. For a particular color, shape and odor state to be actuated, inputs are coupled with a pH value. Researchers should specify whether they would like coupled input-outputs such as use of natural input of saliva at pH 6, or decoupled input-outputs where prepared pH solutions generate output designs where input value can be arbitrary. Utilizing natural inputs at pH 6 to output property ranges that do not correspond to their specific value is possible, but requires researchers to develop new composite structures with the primitives to support this. Currently, the material primitives offer a range of color outputs, shape deformations and states in odor strengths but does not compute. Researchers can utilize the Design Space to integrate fluidic logic with the material primitives, for creating nonlinear feedback loops for computation.

**Multi-Property Outputs**

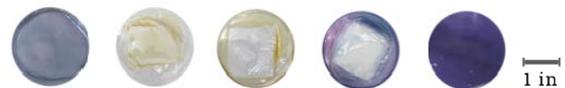

**Figure 12.** Composites with optimal performance (left to right): layering of color-on-odor; the paneling of shape-in-odor, odor-in-shape, odor-in-color; and the mixing of color-odor.

In order to produce multi-property outputs within a single material, we have systematically evaluated the combinability of the three primitives. We have classified three ways the material primitives can be composited: they can be mixed into a homogeneous solution before pouring, layered on top of each other, or synthesized side-by-side so that only the edges mix. We have tested the full combinatory range of composites made with these three methods. Figure 12 depicts the various combinations of primitive composites we determined perform most optimally. A composite was considered successful if it was (i) reactive as intended, (ii) structurally sound, and (iii) reactive within a human-perceptible timescale. These materials exhibit different behaviors when patterned by way of layering, paneling, or mixing the *Organic Primitives* together in various combinations.

**Output Design Parameters**

Designers working with the shape primitive must consider its bimodal pH response. Activating the material with pH 2 will create the most pronounced shape change, while pHs 4 and 10 give more subtle changes. If rate of change is a consideration, more basic pHs result in faster equilibration. The scent-changing material gives a maximum scent response when activated with pH 4, and the scent can be suppressed with pHs 9 or 10. Using pHs greater than 9 to activate the material creates a range of intensities, though perception of these intensities vary between different users. Of the three primitives, the color-changing material had the most linear pH response. pH 2 creates a vivid and instantaneous response. However, the response to very acidic or basic pHs cannot be reversed. If the desired application requires changing color repeatedly, pHs closer to neutral (ie. 6 and 8) should be used.

**Form Factors**

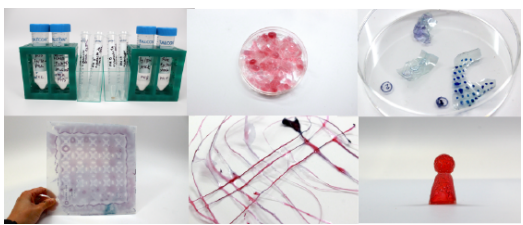

Figure 13. (left to right clockwise) liquid based *Organic Primitives*; odor-color changing gel; color-shape changing films; color changing 3D form; odor-color changing fiber; odor-color-shape changing sheets.

The biopolymer substrates selected for developing the material primitives allows designers to create a range of forms without a drastic alteration in processing techniques. These include fibers, films, sheets, liquids, gels, and solids, as shown in figure 13. pH-reactive color-odor fiber was fabricated using a syringe tube to extrude a formulation of anthocyanin and vanillin within sodium alginate into a calcium chloride solution. The resulting polymer fiber was ionically cross-linked. Thin color-shape changing films and color-odor-shape changing sheets were synthesized by drop-casting molecule-biopolymer solutions atop a sheet of sanded aluminum under a fan. Solid 3D and 2.5D dimensional forms can be made by casting and molding from acrylate or other materials. Primitive subsets can yield diverse properties by creating co-polymers with agar, pectin, gelatin, and other polymers.

**DESIGN SPACE FOR ORGANIC PRIMITIVES**

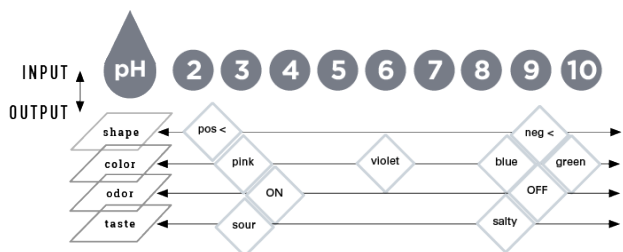

Figure 14. Diagram of pH inputs and outputs from pH 2 - 10.

Here, we showcase examples and techniques for developing more complex systems using the output properties enabled by *Organic Primitives*, as shown in figure 14.

**Animating Properties though Temporal Sequencing**

Sequential and temporal depositions of pH solutions can yield capabilities for physically animating a material's output for color, shape, odor or combinatory changes.

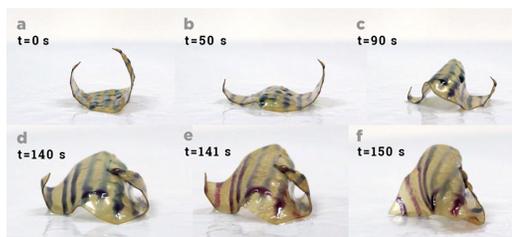

Figure 15. Patterned shape changing primitive activated through temporal sequencing of different pH solutions. Top row was globally activated at pH 8 and bottom row at pH 2.

Beyond sequential and temporal methods, form factor can also enable specific animation capabilities through relative diffusion rates of non-uniform thickness distributions, as shown in figure 16 and figure 15.

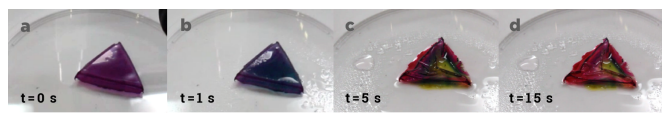

Figure 16. Response of color-odor material molded into triangular 2.5D form to pH 2 solution.

**Encoding Behavior through Patterning**

We describe how multi-property outputs can be achieved through composites and patterning techniques through several examples we have implemented.

*Patterning with Multiple Organic Primitives*
Figure 17 shows a patterned odor-color primitive utilizing hydrophobic fluids such as olive oil for masking particular areas, enabling a globally activated material to possess discrete output patterns. This process termed *cloaking*, can be used to encode hidden information with multiple output properties.

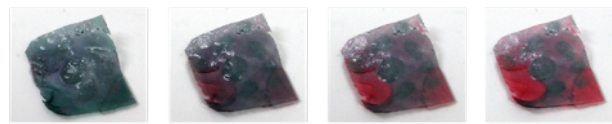

Figure 17. Example of cloaking and disguising information through patterned Odor-Color Organic Primitive.

*Patterning with liquid dopants & pH Solution*
Depending upon how a material is synthesized, fabricated and patterned, activation methods can be interchanged. Instead of using pH solutions as an input to activate the material property changes, they can also be used to pattern and spatially define responsive regions in the material. Figure 15 illustrates how by patterning a shape changing primitive with liquid dopant at pH 10, we can achieve a color-shape changing material.

**Fluidic Logic**

Integrating fluidics with *Organic Primitives* can enable information processing and activation through passive flow or digital control [34] [41] [43]. In this section we highlight different approaches for activation, as shown in figure 11, where different geometry is used to activate and create time-telling and titration devices.

*Ticker for Time-telling*

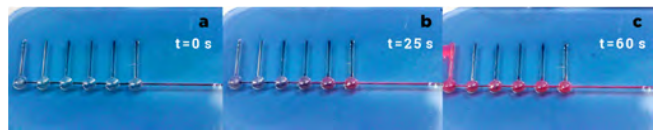

**Figure 18.** Passive diffusion based time-telling device color-odor material composite.

Through diffusion rates of Organic Primitives with pH solutions, we developed a material-based "clock" (figure 18) which changes color to mark specific time increments. The time duration can be tuned by changing the shape, length, or diameter of the channels.

*Gradiator for Titrations*

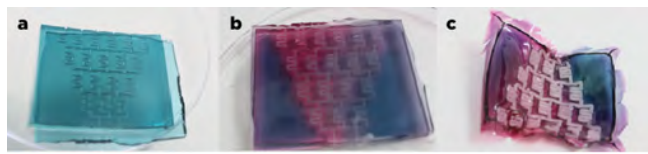

**Figure 19.** Fluidic channels molded then flowed through with pH 2 and pH 10 as inputs to generate a titrated gradient.

Gradiators (figure 19) can be used to titrate solutions by inputting two values, such as pH 2 and pH 10, to create a material with equal values between the endpoints. This can be used to create arrays of flavor, intensities of color, and/or odor strengths.

## APPLICATIONS

Utilizing *Organic Primitives*, we explore molecular design interactions with edibles, cosmetics, environments, and interspecies. The applications shown in this section are working prototypes which illustrate the capabilities of the *Organic Primitives* toolbox. However, in order for them to be functional in the wild, further evaluation will be required. The prototypes were constructed by tuning the concentrations of the sensor-actuators and integrating specific additives.

*Hacking Materiality of Objects*
The information displayed on interfaces about our bodies, food, environment, and organisms are often divorced from the actual systems themselves. By utilizing pre-existing objects as sensor-actuator platforms, we can take advantage of the contextual information they embody and design new relationships with the systems they represent. Figure 20 in the applications section shows an example of how *Organic Primitives* can enable an additional layer of information on objects, transforming an ordinary objects to encode meaning beyond its initial function.

*Sensory Augmentation*
An everyday object can gain significant expressiveness if given the ability to share information about itself through its materiality. The material primitives can be used for sensory augmentations through odor, color, shape, and taste change, where novel outputs can be encoded into every day objects. Figure 21 highlights how odor-changing utensils can be used to simulate a sweet perception of food, without sugar added.

**Edible**
The pH of food varies across a range of pH 2-8. For example soda has a pH of around 2.5 and many fermented foods such as pickles and cheese, range between pH 3-6. Milk and vegetables are more alkaline and range from 6-8. Oral supplements can range from pH 2-10, with vitamin C oral supplement being pH 2.4 and milk of magnesia at pH 10.5. In this section, we explore food-based experiences through the use of *Organic Primitives*.

*Apple Sensor-Display*

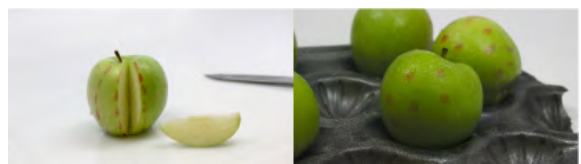

**Figure 20.** Prototype of a patterned edible film coated onto the surface of an apple transforming into a sensor-display. (left) Apple displaying marks showing kids where to cut. (right) Spotted patterns communicate contaminated fruit.

An apple can be transformed into a display by patterning color primitives tuned with methylcellulose MethocelF50 deposited onto edible starch paper. When a user washes the apple, it can display information about itself such as - where to cut for kids; recipe information (ie. required apples for pie); and more advanced application for sensing environmental contaminants during transport for food safety, as illustrated in figure 20.

*Biodegradable Smart Utensils*

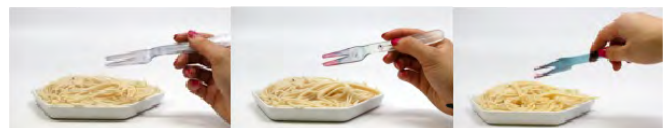

**Figure 21.** Prototype of utensils at varying levels of degradability. (right) Fully edible and biodegradble utensil. (left) Before and after image of dip-coated, 3D-printed fluidic utensil.

We explored multiple methods of creating responsive utensils at different levels of degradability, as shown in figure 21. The first is a fully biodegradable utensil using odor-color changing material primitives that serve to augment the taste and smell of food by releasing a sweet smell instead of adding sugar. The second utensil is partially degradable, where the primitive was dip-coated onto the prong end of a 3D-printed utensil with embedded fluidic channels. The fluidic channels serve as a passive time telling mechanism for pacing the food intake for a user, by changing color.

*Shape Programmable Pasta*
Chitosan biopolymer derived from crustacean shells have been explored as dietary fiber and edible coating the food industry [11]. In figure 22, we prototype a dynamic food which enables chefs to alter shape of the pasta to enable different levels of flavor retention and sensory augmentation in pasta. The shape-changing pasta was made through patterning specific areas to define shape deformations. Other form factors for dynamic food can be used as on-plate communication and embellishments to incorporate condiments as input for simultaneous control of flavor, texture, and appearance. While not implemented in this example, we envision dynamic foods which can change based on user preferences and dietary constraints.

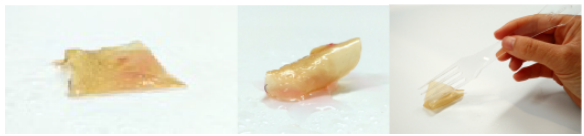

**Figure 22. Prototype of dynamic food that alters in shape based on pH of associated sauces.**

### Body Care and Cosmetics
*Expressive Cosmetic Display*
Figure 23 exhibits how our molecular sensor-actuators can be used as components for cosmetic applications, by integrating primitives for dynamic makeup or for leverage pH data exerted from human body fluids.

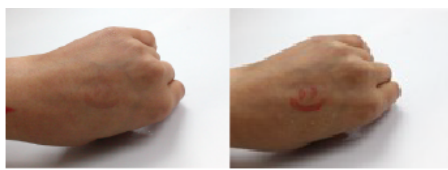

**Figure 23. Interaction prototype of cosmetics displaying hidden information by tuning color-odor primitive with vegetable glycerin.**

Depending upon the area, which they determine to be applied, the material can sense and output based on natural inputs defined by creation of body fluid. For example, applied under the eye directs sensing of tears to change color of eye-makeup. Applied throughout the back can implicate sensing of lactic acid in sweat produced by skin microflora to activate an odor release. Personal care products can also be used as inputs for designing interactions activated by household chemicals such as soaps, typically spanning between pH 8-12. Cosmetics have typically been used to accentuate facial features and communicate character. Dynamic cosmetics can be used to alter human appearances for theater and performance since character progressions tend to change rapidly. Further investigation is needed to bring this into fruition.

*Saliva sensing for oral health*

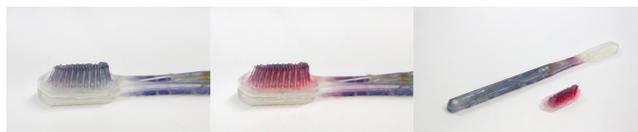

**Figure 24. Toothbrush with modular, interchangeable *Organic Primitive* bristles.**

pH is an important factor for maintaining oral health as certain bacteria contribute to fluctuations. Prolonged periods of acidic oral environment leads to cavities. The oral environment of the mouth maintains an average of pH 6 in the mouth. Figure 24 is an illustrative example of material primitives integrated to create a responsive toothbrush where the bristles can sense and survey the pH of a user's oral environment.

### Interspecies
Plants, microbes, insects and animals utilize chemical signals for communication. In this section, we explore interactions with other species with pH-reactive *Organic Primitives*.

*Augmenting Plant Aesthetics & Fragrance*
Gardens of flowers are often used to enhance the aesthetic environment. Gardeners often monitor and alter the pH of plants to manipulate the way they grow.

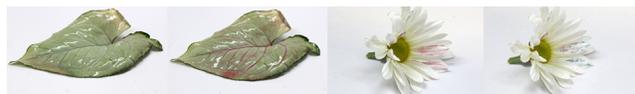

**Figure 25. Odor-color changing *Organic Primitives* coated on the surfaces of leaves (top row) and flower pedals (bottom row).**

Figure 25 shows an implementation of color and smell-changing leaves and flowers. These can be used for household decor, fostering human-plant and human-microbial interactions when users water the plant using different pH solutions.

*Bionic Fruit for Fermentation*

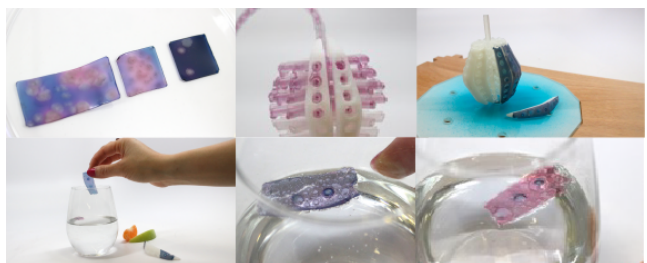

**Figure 26. Prototype of Fructus, a bionic fruit with a fluidic 3D printed core which electromechanically deposits edible pH-reactive sensor-actuators onto its surface.**

Fructus is a bionic fruit prototype which generates skins of dynamic edible films, pictured in figure 26. A microcontroller with syringes of the three primitives are set in the table below. The device deposits the materials into the 3D printed fluidic core of the bionic fruit, through a syringe pump mechanism driven by a stepper motor. The outlet of the fluidics draws the materials to coat the surfaces of the "fruit" slices. When dried, users can peel off to use as condiment, flavoring, embellishment, or integrate in non-food applications. Future work can offer these as tools for chefs to communicate to their guests by remotely triggering Fructus to synthesize different flavors, textures, and tastes in the form of dynamic edible films.

### Environment
In natural systems, chemical exchanges represent the communication pathways between micro- and macro-scale ecosystems. On a macro scale, pH values can be an indicator of environmental issues from ocean acidification to acid rain. In this section, we explore how to enable users to form more engaged relationships with the environment by using rain as an input and carrier of environmental information for sensory augmentation and display.

*Bleeding Umbrella*
Fossil fuel combustion from automobile and industrial manufacturing generates emissions of sulfur dioxide and nitrogen oxide which contribute to rainfall acidity and environmental pollution. We developed a bleeding umbrella prototype shown in figure (figure 27), using *Organic Primitives* to sense the rain and display red streaks on the surface "as if the sky was bleeding", in the presence of acid rain. This provides the user with visceral information from the sky, creating a contextual connection between the contents of rain water and environmental contaminants.

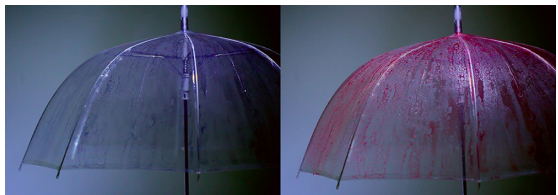

Figure 27. Before and after image of umbrella utilizing *Organic Primitives* for acid rain sensing and display.

*Architectural Material to Augment Smell of the Rain*

The smell of rain on a rural landscape is often romanticized by poets and artists. Within an urban environment, the smell of the rain is a byproduct of hydration of materials such as asphalt and concrete. In figure 28, we created an architectural material which outputs odor and color by augmenting the smell of different places. These materials can give architects an additional sensory medium and experience to design with.

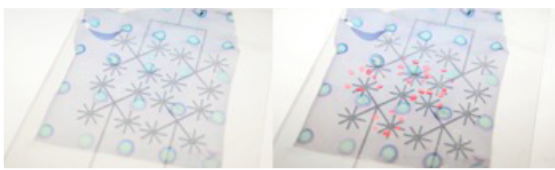

Figure 28. Odor and color changing architectural material for rain augmentation.

## DISCUSSION AND FUTURE WORK

Our use of the term *organic* is rooted in the carbon-based compounds that our sensor-actuator materials are built from, as distinguished from non-flat displays in organic user interfaces (OUI). Although the material primitives do not require a computer or wires to operate, they serve as sensors and actuators which offer researchers a method for interfacing with information within fluids. This approach of synthesizing color, odor and shape changing material sensor-actuators enables a variety of advantages and design opportunities as these systems: (i) have self-contained functionality as a sensor, actuator and energy source; (ii) can be manifested in different form factors and states of matter, from liquid to solid to vapor; (iii) can be integrated into both biological and electronic systems; (iv) are biocompatible, biodegradable, and edible; (v) are compact, soft, muted, and unobtrusive; (vi) open up additional modalities including taste and smell. Through this research, we hope to encourage other work in this non-traditional HCI landscape.

### Offloading Design Parameters to Software Tools

Data generated from our material characterization and evaluation serve as our design parameters in our material system. This nonlinear and multi-dimensional data can be integrated into software design tools which incorporate specific material formulations, recipes and fabrication conditions as digital and material design processes. Furthermore, the synthesis and printing of these materials can be automated onto printing platforms.

### Expanding the library of Organic Primitives

While pH serves as a starting point in this body of work, we envision a library of material primitives which can facilitate designers to interact with a variety of organic systems, across a range of signals. Primitives which can transduce molecular signals such as glucose, eH, $pCO_2$, sodium, magnesium, calcium, potassium, hormones and dioxins, can further enrich possibilities for human scale interaction with organic systems. The methods introduced in this paper can be used to develop sensor-actuators that are responsive to any number of chemical and biological stimuli. As the library of primitives expand, material robotics which can sense and output a multiplicity of behaviors can be possible. This can be achieved by utilizing the introduced techniques from the design space with multiple molecular I/O dopants, to build complex logic and behavior.

## CONCLUSION

*Organic Primitives* offer designers a toolbox to create multimodal actuation to "hack" the materiality of everyday objects - transforming them into both sensors and information displays. Because of the biocompatible nature of our sensor-actuators, we are able to utilize fluids from raindrops, ocean water, body fluids, food sources, and plant/microbe excretions as natural inputs to activate the output properties of the primitives. Our approach of employing organic compounds as molecular I/O offers a method to leverage an object's contextual meaning to enable novel sensory experiences by augmenting its material properties.

## ACKNOWLEDGMENTS

This paper was made possible by the support of Neri Oxman, Joi Ito, Linda Peterson, Jake Bernstein, Ivan Sysoev, and members of the Mediated Matter Group. We thank Laia Mogas, Judith Amores, Dhruv Jain, Yujie Hong, Chikara Inamura, Hershel Macaulay, Jifei Ou, Udayan Umapathi, Philipp Schoessler, Felix Heibeck, Basheer Tome, Nick Barry and Xiao Xiao for their insight and ideation. We are grateful for the generous feedback provided by David Kong from MIT Lincoln Laboratory, Niels Holten-Andersen from MIT Materials Science and Engineering, Manu Prakash from Stanford University and Pia Sorensen from Harvard University, at different stages of this research. We appreciate the tools provided by the MIT Center for Bits and Atoms and Ike Feitler for the spectrophotometer.